\documentclass[a4paper, 11pt]{article}
\usepackage[table]{xcolor}
\usepackage{setspace}
\usepackage{bbm}
\usepackage{hyperref}
\definecolor{mycolor}{RGB}{0,88,204}
\definecolor{mydarkcolor}{RGB}{128, 0, 32}
\hypersetup{
  colorlinks=true,
  linkcolor=mycolor,
  urlcolor=mycolor,
  citecolor=mycolor
}
\usepackage{amsmath}
\usepackage{geometry}

\usepackage{fancyhdr}

\fancypagestyle{plain}{%
  \fancyhf{}
  \fancyhead[R]{\thepage}
  
}

\usepackage{mathptmx}

\usepackage{titlesec}

\titleformat{\section}
  {\normalfont\large\centering}{\thesection.}{1em}{\MakeUppercase}
\titleformat{\subsection}
  {\normalfont\centering}{\thesubsection.}{1em}{\MakeUppercase}

  \titlespacing{\paragraph}{10pt}{0pt}{6pt}[0pt]

\usepackage{svg}
\usepackage{tikz-cd} 			
\usepackage{tikz}
\usetikzlibrary{arrows.meta}
\usetikzlibrary{calc}
\usetikzlibrary{positioning}
\usepackage{listings}
\usepackage[T1]{fontenc}
\usepackage[utf8]{inputenc}
\usepackage{amssymb}
\usepackage{chngcntr}
\definecolor{keywordcolor}{rgb}{0.7, 0.1, 0.1}   
\definecolor{tacticcolor}{rgb}{0.0, 0.1, 0.6}    
\definecolor{commentcolor}{rgb}{0.4, 0.4, 0.4}   
\definecolor{symbolcolor}{rgb}{0.0, 0.1, 0.6}    
\definecolor{sortcolor}{rgb}{0.1, 0.5, 0.1}      
\definecolor{attributecolor}{rgb}{0.7, 0.1, 0.1} 

\lstnewenvironment{code}[1][]%
{
   \noindent\newline
   \minipage{\linewidth}
   \vspace{0.5\baselineskip}
   \lstset{
 	frame = lrtb,
 	rulecolor=\color{mycolor},
 	escapeinside={/*!}{!*/},
	language=lean,
	aboveskip = 5mm,
	belowskip = 5mm,
	xleftmargin=2mm,
	xrightmargin=2mm,
	}
	}
{\endminipage\newline}
\lstnewenvironment{codeLong}[1][]%
{
   \lstset{
 	frame = lrtb,
 	rulecolor=\color{mycolor},
 	escapeinside={/*!}{!*/},
	language=lean,
	aboveskip = 5mm,
	belowskip = 5mm,
	xleftmargin=2mm,
	xrightmargin=2mm,
	}
	}
{}
\usepackage{xpatch}
\usepackage{listings}
\usepackage{realboxes}
\definecolor{mycolorSubtle}{RGB}{245,250,255}
\DeclareRobustCommand{\myinline}{\lstinline}
\makeatletter
\xpretocmd\myinline{\Colorbox{mycolorSubtle}\bgroup\appto\lst@DeInit{\egroup}}{}{}
\makeatother

\lstdefinestyle{mystyle}
{
  language=lean,
  backgroundcolor=\color{gray},
}
\newcommand{\codeLink}[1]{
  \vspace{-0.5cm}\hfill\href{https://github.com/HEPLean/HepLean/blob/1b951994ae3d882242b02d23957ef1ee7fa05f3d/HepLean/#1}{(source)}
  }
  
 \newcommand{\textLink}[1]{\href{https://github.com/HEPLean/HepLean/blob/1b951994ae3d882242b02d23957ef1ee7fa05f3d/HepLean/#1}{source}}
 \newcommand{\textLinkB}[1]{\href{https://github.com/HEPLean/HepLean/blob/1b951994ae3d882242b02d23957ef1ee7fa05f3d/HepLean/#1}{(source)}}

\newcommand{\syntaxElab}[2]{ 
  \arrayrulecolor{mycolor}
  \begin{center}
    \begin{tabular}{|p{1.7in} | p{4in}|}
    \hline
    \hfill {#1} & {#2} \\
    \hline
    \end{tabular}
    \end{center}
  \arrayrulecolor{black}
}

\newcommand{\proofstep}[3]{
  \arrayrulecolor{mycolor}
\begin{center}
\begin{tabular}{|p{3in}| p{3in}|}
\hline
{#1
}\newline 
\hrule~\newline
#2
  & ~\newline
\makebox[3in]{%
#3}
  \\ \hline
\end{tabular}
\end{center}
\arrayrulecolor{black}
}

\newcommand{\tensorTree}[1]{
\begin{center}
  \fcolorbox{mycolor}{white}{%
#1}
\end{center}
}

\title{Formalization of physics index notation in Lean 4}
\author{Joseph Tooby-Smith \\ \textit{Reykjavik University, Reykjavik, Iceland}}
\date{\today}

\begin{document}
\counterwithin{lstlisting}{section}
\maketitle
\vspace{-1cm}
\begin{abstract}
The physics community relies on index notation to effectively manipulate types of tensors.
This paper introduces the first formally verified implementation of index notation in the
interactive theorem prover Lean 4. By integrating index notation into Lean, we bridge the gap between 
traditional physics notation and formal verification tools, 
making it more accessible for physicists to write and prove results within Lean.
We also open up a new avenue through which AI tools can be used to prove results
related to tensors in physics.
Behind the scenes our implementation leverages a novel application of category theory.
\end{abstract}

\section{Introduction}

In previous work~\cite{HepLean}, the author initiated the digitalization (or formalization) 
of high energy physics 
results using the interactive theorem prover Lean 4~\cite{lean} in a project called HepLean. 
Lean is a
programming language with syntax resembling traditional pen-and-paper mathematics. 
Users can write definitions, theorems, and proofs in Lean,
which are then automatically checked for correctness using dependent-type theory.
The HepLean project is driven by four primary motivations: (1) to facilitate easier look-up of results
 through a linear storage of information; (2) to support the creation and proof of new results using 
 automated tactics and AI tools; (3) to facilitate checking of result in high energy physics for correctness; 
 and (4) to introduce new pedagogical methods for high-energy physics and computer science.

HepLean is part of a broader movement of projects
to formalize parts, or all, of 
mathematics and science. The largest of these projects is Mathlib~\cite{mathlib}, which aims to formalize
mathematics. Indeed, HepLean uses many results from Mathlib, and has Mathlib as an import.
Other projects in mathematics include the ongoing effort led by Kevin Buzzard to formalize the proof of Fermat's
Last Theorem into Lean~\cite{FLT}. 
In the realm of the sciences, Bobbin et al.~\cite{josephson} looks at absorption theory, thermodynamics, and kinematics in Lean, 
whilst the package SciLean~\cite{SciLean}, is a push in the direction of scientific computing within Lean. 

Physicists rely heavily on specialized notation to express mathematical concepts succinctly. 
Among these index notation is particularly prevalent,
 as it provides a compact and readable way to represent specific types of tensors and operations
 between them. Such tensors form a backbone of modern physics. 

Having a way to use index notation in Lean is crucial for the digitalisation of 
results from high energy physics. 
In addition to making results from high energy physics easier to write and prove in Lean, 
it will make the syntax more familiar to high energy physicists.
However, there are challenges in implementing index notation in Lean,
namely, the need for a formal and rigorous implementation that is also easy and nice to use. 
Such an implementation can now be found as part of HepLean:
\begin{center}
\url{https://heplean.github.io/HepLean/}
\end{center}
and is the subject of this paper.
We hope that the implementation presented here will
not only enhance usability of Lean but also promotes the adoption of formal methods in the 
physics community.

To give an example of the up-shot of our implementation, the result regarding Pauli matrices
that $\sigma^{\nu \alpha \dot \beta}\sigma_\nu^{ \alpha' \dot \beta'} = 2 \epsilon^{\alpha \alpha'}\epsilon^{\beta \beta'}$ 
is written in Lean, using our implementation, as follows:
\begin{code}
{pauliCo | ν α β ⊗ pauliContr | ν α' β' = 2 •ₜ εL | α α' ⊗ εR | β β'}ᵀ
\end{code}
Lean will correctly interpret this result as a tensor expression with the correct contraction of indices 
and permutation of indices between each side of the expression.
Our implementation can  handle different species of tensors, 
for example real Lorentz tensors, complex Lorentz tensors and ordinary tensors (e.g., vectors and matrices).
That said, at the time of writing only the most involved, complex Lorentz tensors, have been implemented.

Previous implementations of index notation have been made in programming languages like Haskell~\cite{haskellPaper}. However, the programs they appear in do not
 provide the formal verification capabilities inherent in Lean. 
 The formal verification requirement of Lean, and the need to cover different species of tensors, introduces unique challenges in implementing index 
 notation, necessitating (what we believe is) a novel solution.

In Section~\ref{sec:Implementation} of this paper, we will discuss the details of our implementation. 
In Section~\ref{sec:examples} we will give two examples of definitions, theorems and proofs in Lean using index notation 
to give the reader an idea of how our implementation works in practice.
The first of these examples involves a lemma regarding the contraction of indices of symmetric and antisymmetric tensors. 
The second involves examples related to the Pauli matrices and bispinors. 
We finish this paper in Section~\ref{sec:future} by discussing potential future work related to this project.

\subsection*{Notation}
In this paper, we will follow Lean's notation for types and terms. 
For example if \myinline|C| is a type (which can be thought of as similar to a set),
then \myinline|c : C| says that \myinline|c| is an element of the type \myinline|C|. 
In addition, instead of having two streams of notation for mathematical objects, one from the Lean 
code and one in LaTeX, we will use Lean code throughout to represent mathematical objects.
Throughout Section~\ref{sec:Implementation}, we will assume a basic knowledge of the theory of 
symmetric monoidal categories.
\section{Implementation of index notation into Lean 4}\label{sec:Implementation}

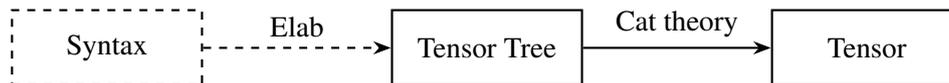
\begin{figure}
  \centering
  \begin{tikzpicture}[
    thick,
    >=Stealth,
    box/.style={rectangle, draw, minimum height=1cm, minimum width=2.5cm, align=center},
    dashedbox/.style={rectangle, draw, dashed, minimum height=1cm, minimum width=2.5cm, align=center}
  ]
  
  \node[dashedbox] (syntax) at (0,0) {Syntax};
  \node[box] (tensorTree) at (5,0) {Tensor Tree};
  \node[box] (tensor) at (10,0) {Tensor};
  
  \draw[->, dashed] (syntax) -- node[above] {Elab} (tensorTree);
  \draw[->] (tensorTree) -- node[above] {Cat theory} (tensor);
  
  \end{tikzpicture}
  \caption{Overview of the implementation of index notation in Lean. The 
  solid lines represent formally verified parts of the implementation.}
  \label{fig:overviewFlow}
\end{figure}

Our implementation of index notation can be thought of as three different representations of tensor 
expressions and maps between them. This is illustrated in Figure~\ref{fig:overviewFlow}.

The first representation is \emph{syntax}. This can (roughly) be thought of as the informal string
that represents the tensor expression. It is what the user interacts with when 
writing results in Lean, and what appears in raw Lean files. 
An example of syntax is given in the code snippet we gave above 
for Pauli matrices.\footnote{In practice there is a representation before syntax, 
 which is as a sequence of tokens in file which is parsed by Lean into a more structured syntax. For simplicity,
 we think of these representations as the same here.} 

The second representation of a tensor expression is a \emph{tensor tree}. This representation is 
mathematically formal, but easy to use and manipulate. 
It is a structured tree which has different types 
of nodes for each of the main operations that one can perform on tensors.

The process of going from syntax to a tensor tree is done via an elaborator 
which follows a number of (informally defined) rules.

The third and final representation is a bona-fide \emph{tensor}. This representation 
is the mathematical object that we are actually interested in. However, in going to this representation
we lose the structure of the tensor expression itself, making it difficult to work with. 

The process of going from a tensor tree to a tensor involves properties from a symmetric monoidal 
category of representations. 

Throughout our discussion we will need the notion of a `tensor species'.
Thus, we first give formal definition of this, before discussing each of the above 
representations and the processes between them in more detail.

\subsection{Tensor Species}

A tensor species is a novel formalization of the data needed to define, e.g., complex Lorentz tensors, 
real Lorentz tensors or ordinary tensors (e.g., vectors and matrices). The word `species' is a nod to `graphical species'
defined in~\cite{JOYAL2011105,raynor2021graphical}, from which our construction is inspired.

We will start by giving the complete definition of a tensor species in Lean, and then 
dissect this definition, discussing each of the components in turn. 

In Lean a tensor species is defined as follows:
\begin{codeLong}
/-- The structure of a type of tensors e.g., Lorentz tensors, ordinary tensors 
  (vectors and matrices), complex Lorentz tensors. -/
structure TensorSpecies where
  /-- The commutative ring  over which we want to consider the tensors to live in,
    usually `ℝ` or `ℂ`. -/
  k : Type
  /-- An instance of `k` as a commutative ring. -/
  k_commRing : CommRing k
  /-- The symmetry group acting on these tensor e.g. the Lorentz group or SL(2,ℂ). -/
  G : Type
  /-- An instance of `G` as a group. -/
  G_group : Group G
  /-- The colors of indices e.g. up or down. -/
  C : Type
  /-- A functor from `C` to `Rep k G` giving our building block representations.
    Equivalently a function `C → Rep k G`. -/
  FD : Discrete C ⥤ Rep k G
  /-- A specification of the dimension of each color in C. This will be used for explicit
    evaluation of tensors. -/
  repDim : C → ℕ
  /-- repDim is not zero for any color. This allows casting of `ℕ` to `Fin (S.repDim c)`. -/
  repDim_neZero (c : C) : NeZero (repDim c)
  /-- A basis for each Module, determined by the evaluation map. -/
  basis : (c : C) → Basis (Fin (repDim c)) k (FD.obj (Discrete.mk c)).V
  /-- A map from `C` to `C`. An involution. -/
  τ : C → C
  /-- The condition that `τ` is an involution. -/
  τ_involution : Function.Involutive τ
  /-- The natural transformation describing contraction. -/
  contr : OverColor.Discrete.pairτ FD τ ⟶ 𝟙_ (Discrete C ⥤ Rep k G)
  /-- Contraction is symmetric with respect to duals. -/
  contr_tmul_symm (c : C) (x : FD.obj (Discrete.mk c))
      (y : FD.obj (Discrete.mk (τ c))) :
    (contr.app (Discrete.mk c)).hom (x ⊗ₜ[k] y) = (contr.app (Discrete.mk (τ c))).hom
    (y ⊗ₜ (FD.map (Discrete.eqToHom (τ_involution c).symm)).hom x)
  /-- The natural transformation describing the unit. -/
  unit : 𝟙_ (Discrete C ⥤ Rep k G) ⟶ OverColor.Discrete.τPair FD τ
  /-- The unit is symmetric. -/
  unit_symm (c : C) :
    ((unit.app (Discrete.mk c)).hom (1 : k)) =
    ((FD.obj (Discrete.mk (τ (c)))) ◁
      (FD.map (Discrete.eqToHom (τ_involution c)))).hom
    ((β_ (FD.obj (Discrete.mk (τ (τ c)))) (FD.obj (Discrete.mk (τ (c))))).hom.hom
    ((unit.app (Discrete.mk (τ c))).hom (1 : k)))
  /-- Contraction with the unit does nothing. -/
  contr_unit (c : C) (x : FD.obj (Discrete.mk (c))) :
    (λ_ (FD.obj (Discrete.mk (c)))).hom.hom
    (((contr.app (Discrete.mk c)) ▷ (FD.obj (Discrete.mk (c)))).hom
    ((α_ _ _ (FD.obj (Discrete.mk (c)))).inv.hom
    (x ⊗ₜ[k] (unit.app (Discrete.mk c)).hom (1 : k)))) = x
  /-- The natural transformation describing the metric. -/
  metric : 𝟙_ (Discrete C ⥤ Rep k G) ⟶ OverColor.Discrete.pair FD
  /-- On contracting metrics we get back the unit. -/
  contr_metric (c : C) :
    (β_ (FD.obj (Discrete.mk c)) (FD.obj (Discrete.mk (τ c)))).hom.hom
    (((FD.obj (Discrete.mk c)) ◁ (λ_ (FD.obj (Discrete.mk (τ c)))).hom).hom
    (((FD.obj (Discrete.mk c)) ◁ ((contr.app (Discrete.mk c)) ▷
    (FD.obj (Discrete.mk (τ c))))).hom
    (((FD.obj (Discrete.mk c)) ◁ (α_ (FD.obj (Discrete.mk (c)))
      (FD.obj (Discrete.mk (τ c))) (FD.obj (Discrete.mk (τ c)))).inv).hom
    ((α_ (FD.obj (Discrete.mk (c))) (FD.obj (Discrete.mk (c)))
      (FD.obj (Discrete.mk (τ c)) ⊗ FD.obj (Discrete.mk (τ c)))).hom.hom
    ((metric.app (Discrete.mk c)).hom (1 : k) ⊗ₜ[k]
      (metric.app (Discrete.mk (τ c))).hom (1 : k))))))
    = (unit.app (Discrete.mk c)).hom (1 : k)
\end{codeLong}

Let us work through this definition piece by piece.
The first part of the definition of a tensor species is a type \myinline|k| 
which is a commutative ring via \myinline|k_commRing|. 
For a given \myinline|S : TensorSpecies|, we retrieve \myinline|k| and all of the 
fields of \myinline|TensorSpecies| via \myinline|S.k| etc. 
For the tensor species of complex Lorentz tensors, denoted 
\myinline|complexLorentzTensor|, the ring \myinline|complexLorentzTensor.k| is the ring of complex numbers.
In general, \myinline|k| will be the ring over which the tensors are defined.

The second part of the definition of a tensor species is a type \myinline|G| 
which is a group via \myinline|G_group|. For complex Lorentz tensors, 
\myinline|complexLorentzTensor.G| is the group $SL(2, \mathbb{C})$, 
whilst for real Lorentz tensors it will be the Lorentz group.
It is the group which acts on our tensors.

The next part of the definition is a type \myinline|C| which we call the type of colors.
For complex Lorentz tensors, \myinline|complexLorentzTensor.C| is equal to the type:
\begin{code}
inductive Color
  | upL : Color
  | downL : Color
  | upR : Color
  | downR : Color
  | up : Color
  | down : Color
\end{code}
which contains six colors.
Colors can be thought of as labels for each of the building 
block representations making up the tensor species.  This is made formal by the
next part of the definition of a tensor species, a functor 
\myinline|FD| from the discrete category formed by \myinline|C|, \myinline|Discrete C|,to 
the category of representations of \myinline|G| over \myinline|k|, \myinline|Rep k G|.
The category  \myinline|Rep k G| and its properties are defined in Mathlib, along 
with the necessary category theory we use in this paper.

\sloppy The functor \myinline|FD| assigns to each color a representation of \myinline|G| over \myinline|k|.
Note that for \myinline|c : C| to apply \myinline|FD| we have to write 
\myinline|FD.obj (Discrete.mk c)| in Lean. This is somewhat cumbersome, hence in what 
follows we will abbreviate this to \myinline|FD c|.
For complex Lorentz tensors, the functor \myinline|complexLorentzTensor.FD| is defined as:
\begin{code}
FD := Discrete.functor fun c =>
  match c with
  | Color.upL => Fermion.leftHanded
  | Color.downL => Fermion.altLeftHanded
  | Color.upR => Fermion.rightHanded
  | Color.downR => Fermion.altRightHanded
  | Color.up => Lorentz.complexContr
  | Color.down => Lorentz.complexCo
\end{code}
The representations appearing here are: 
\begin{itemize}
  \item The representation of left-handed Weyl fermions, 
    denoted in Lean as \myinline|Fermion.leftHanded|, and corresponding to the 
    representation of $SL(2, \mathbb{C})$ taking $v \mapsto M v$ for $M \in SL(2, \mathbb{C})$.
  \item \sloppy The representation of `alternative' left-handed Weyl fermions, 
    denoted in Lean as \myinline|Fermion.altLeftHanded|, and corresponding to the 
    representation of $SL(2, \mathbb{C})$ taking $v \mapsto M^{-1 T} v$ for $M \in SL(2, \mathbb{C})$.
  \item The representation of right-handed Weyl fermions, 
    denoted in Lean as \myinline|Fermion.rightHanded|, and corresponding to the 
    representation of $SL(2, \mathbb{C})$ taking $v \mapsto M^\star v$ for $M \in SL(2, \mathbb{C})$.
  \item The representation of `alternative' right-handed Weyl fermions,
    denoted in Lean as \myinline|Fermion.altRightHanded|, and corresponding to the 
    representation of $SL(2, \mathbb{C})$ taking $v \mapsto M^{-1 \dagger} v$ for $M \in SL(2, \mathbb{C})$.
  \item The representation of contravariant Lorentz tensors, 
    denoted in Lean as \myinline|Lorentz.complexContr|, and corresponding to the 
    representation of $SL(2, \mathbb{C})$ induced by the homomorphism of $SL(2, \mathbb{C})$ into 
    the Lorentz group and the contravariant action of the Lorentz group on four-vectors.
  \item The representation of covariant Lorentz tensors,
     denoted in Lean as \myinline|Lorentz.complexCo|, and corresponding to the 
    representation of $SL(2, \mathbb{C})$ induced by the homomorphism of $SL(2, \mathbb{C})$ into 
    the Lorentz group and the covariant action of the Lorentz group on four-vectors.
\end{itemize}
As an example of how these are defined in Lean, the representation of left-handed Weyl fermions  \myinline|Fermion.leftHanded|
is given by:
\begin{codeLong}
/-- The vector space ℂ^2 carrying the fundamental representation of SL(2,C).
  In index notation corresponds to a Weyl fermion with indices ψ^a. -/
def leftHanded : Rep ℂ SL(2,ℂ) := Rep.of {
  /- The function from SL(2,ℂ) to endomorphisms of LeftHandedModule 
    (which corresponds to the vector space ℂ^2). -/
  toFun := fun M => {
    /- Start of the definition of the linear map. -/
    /- The function underlying the linear map. Defined as the dot product. -/
    toFun := fun (ψ : LeftHandedModule) =>
      LeftHandedModule.toFin2ℂEquiv.symm (M.1 *ᵥ ψ.toFin2ℂ),
    /- Proof that the function is linear with respect to addition. -/
    map_add' := by
      intro ψ ψ'
      simp [mulVec_add]
    /- Proof that the function is linear with respect to scalar multiplication. -/
    map_smul' := by
      intro r ψ
      simp [mulVec_smul]
    /- End of the definition of the linear map. -/}
  /- Proof that (the outer) toFun gives the identity map on the identity of SL(2,ℂ). -/
  map_one' := by
    ext i
    simp
  /- Proof that the action of the product of two elements is 
    the product of the actions of the elements. -/
  map_mul' := fun M N => by
    simp only [SpecialLinearGroup.coe_mul]
    ext1 x
    simp only [LinearMap.coe_mk, AddHom.coe_mk, LinearMap.mul_apply, LinearEquiv.apply_symm_apply,
      mulVec_mulVec]}
\end{codeLong} 
We have added some explanatory comments to this code, not seen in the actual Lean code, to give 
the reader an idea of what each part does. Note that the \myinline|Fermion.| part of the name 
of \myinline|Fermion.leftHanded| is dropped in this definition, as it is inherited from 
the Lean namespace in which the definition is made.

The next part of the definition of a tensor species is the map \myinline|repDim|, which assigns to each color a natural number
corresponding to the dimension of the representation associated to that color. The condition is placed on the representations that they are non-empty, i.e., 
that the dimension is not equal to zero \myinline|repDim_neZero|.  
The \myinline|basis| part of the definition of a tensor species gives a basis indexed by \myinline|Fin (repDim c)| 
(numbers from \myinline|0| to \myinline|repDim c - 1|)  of each representation for each \myinline|c : C|.  
We will use this basis in the definition of evaluation of tensor indices. For complex Lorentz tensors, 
these are the standard basis for Lorentz vectors and Weyl-fermions. 

Next in the definition of a  tensor species is the map 
\myinline|τ|, which is an involution via \myinline|τ_involution|.
This assigns to each color its `dual' corresponding to the color it can be contracted with. So, for complex Lorentz tensors 
the map \myinline|τ| is given by: 
\begin{code} 
τ := fun c =>
  match c with
  | Color.upL => Color.downL
  | Color.downL => Color.upL
  | Color.upR => Color.downR
  | Color.downR => Color.upR
  | Color.up => Color.down
  | Color.down => Color.up
\end{code}

The contraction itself is defined in a tensor species by \myinline|contr|, which is a natural transformation from the functor 
\myinline|FD _ ⊗ FD (τ _)|, to the constant functor \myinline|𝟙_ (Discrete C ⥤ Rep k G)| which takes every object in \myinline|C| to the trivial representation 
\myinline|𝟙_ (Rep k G)|. 
This natural transformation is simply the assignment to each color \myinline|c| a linear-map from 
\myinline|FD c ⊗ FD (τ c)| to \myinline|k| which is equivariant with respect to the group action. 
This contraction cannot be defined arbitrarily, but must satisfy the symmetry condition \myinline|contr_tmul_symm|, 
which corresponds to the commutative diagram: 
\begin{equation*}
  \begin{tikzpicture}[>=stealth, node distance=5cm]
    \node (A) at (0, 0) {\lstinline|FD c ⊗ FD (τ c)|};
    \node (B) at (7, 0) {\lstinline|FD (τ c) ⊗ FD c|};
    \node (C) at (0, -2){\lstinline|𝟙_(Rep k G)|};
    \node (D) at (7, -2 ) {\lstinline|FD (τ c) ⊗ FD (τ (τ c))|};
  
    \draw[->] (A) -- (B);
    \draw[->] (A) -- (C);
    \draw[->] (B) -- (D);
    \draw[->] (D) -- (C);

    \node at ($(A)!0.5!(B)$) [above] {\lstinline|β_|};
    \node at ($(A)!0.5!(C)$) [left] {\lstinline|contr c|};
    \node at ($(B)!0.5!(D)$) [right] {\lstinline|𝟙 ⊗ FD.map (e c)|};
    \node at ($(D)!0.5!(C)$) [below] {\lstinline|contr (τ c)|};
  \end{tikzpicture}
\end{equation*}
where \myinline|β_| is the braiding of the symmetric monoidal category, and \myinline|e c| is shorthand for the isomorphism
in \myinline|Discrete C| between \myinline|c| and \myinline|τ (τ c)|, which exists since \myinline|τ| is an involution.

\sloppy Along with the contraction, the definition of a tensor species includes the \myinline|unit|, along with its own symmetry condition 
\myinline|unit_symm|. The unit is a natural transformation from the functor \myinline|𝟙_ (Discrete C ⥤ Rep k G)| to 
\myinline|FD (τ _) ⊗ FD _|. This is the assignment to each color \myinline|c| an object of 
\myinline|FD (τ c) ⊗ FD c| which is invariant with respect to the group action. 
The symmetry condition \myinline|unit_symm| is represented by the commutative diagram:
\begin{equation*}
  \begin{tikzpicture}[>=stealth, node distance=5cm]
    \node (A) at (0, 0) {\lstinline|𝟙_(Rep k G)|};
    \node (B) at (7, 0) {\lstinline|FD (τ c) ⊗ FD c|};
    \node (C) at (0, -2){\lstinline|FD (τ (τ c)) ⊗ FD (τ c)|};
    \node (D) at (7, -2 ) {\lstinline|FD (τ c) ⊗ FD (τ (τ c))|};
  
    \draw[->] (A) -- (B);
    \draw[->] (A) -- (C);
    \draw[->] (D) -- (B);
    \draw[->] (C) -- (D);

    \node at ($(A)!0.5!(B)$) [above] {\lstinline|unit c|};
    \node at ($(A)!0.5!(C)$) [left] {\lstinline|unit (τ c)|};
    \node at ($(D)!0.5!(B)$) [right] {\lstinline|𝟙 ⊗ FD.map (e c)|};
    \node at ($(C)!0.5!(D)$) [below] {\lstinline|β_|};
  \end{tikzpicture}
\end{equation*}
The next condition, \myinline|contr_unit|, makes formal the statement that contraction with the 
unit does nothing. It corresponds to the diagram: 
\begin{equation*}
  \begin{tikzpicture}[>=stealth, node distance=2cm and 2.5cm] 
    \node (A) at (0, 0) {\lstinline|FD c|};
    \node[right=of A] (E) {\lstinline|FD c ⊗ 𝟙_(Rep k G)|};
    \node[right=of E] (B) {\lstinline|FD c ⊗ (FD (τ c) ⊗ FD c)|};
    \node[below=of A] (C) {\lstinline|𝟙_(Rep k G) ⊗ FD c|};
    \node[below=of B] (D) {\lstinline|(FD c ⊗ FD (τ c)) ⊗ FD c|};
  
    \draw[->] (A) -- (E);
    \draw[->] (E) -- (B);
    \draw[->] (C) -- (A);
    \draw[->] (B) -- (D);
    \draw[->] (D) -- (C);

    \node at ($(A)!0.4!(E)$) [above] {\lstinline|(ρ_)⁻¹|};
    \node at ($(E)!0.45!(B)$) [above] {\lstinline|𝟙 ⊗ unit c|}; 
    \node at ($(A)!0.5!(C)$) [left] {\lstinline|λ_|};
    \node at ($(D)!0.5!(B)$) [right] {\lstinline|(α_)⁻¹|};
    \node at ($(C)!0.5!(D)$) [below] {\lstinline|contr c ⊗ 𝟙|};
  \end{tikzpicture}
\end{equation*}
where \myinline|ρ_| is the right-unitor, \myinline|λ_| is the left-unitor, and \myinline|α_| is the associator
in the category \lstinline|Rep k G|. 

The final part of the definition of a tensor species is the metric, \myinline|metric|,
and its interaction, \myinline|contr_metric|, with the contraction and unit. 
The metric is a natural transformation from the functor \myinline|𝟙_ (Discrete C ⥤ Rep k G)|
to the functor \myinline|FD _ ⊗ FD _|. It thus represents the assignment to each color \myinline|c| an object of
\myinline|FD c ⊗ FD c| which is invariant with respect to the group action.
The metric can be used to change an index into a dual index. The condition \myinline|contr_metric|
corresponds to the diagram:
\begin{equation*}
  \begin{tikzpicture}[>=stealth, node distance=1.5cm and 1cm] 
    \node (B1) {\myinline|𝟙_(Rep k G)|};
    \node[left=of B1] (A1) {\lstinline|𝟙_(Rep k G) ⊗ 𝟙_(Rep k G)|};
    \node[below=of A1] (A2) {\lstinline|(FD c ⊗ FD c) ⊗ (FD (τ c) ⊗ FD (τ c))|};
    \node[below=of A2] (A3) {\lstinline|FD c ⊗ (FD c ⊗ (FD (τ c) ⊗ FD (τ c)))|};
    \node[below=of B1] (B2) {};
    \node[below=of B2] (B3) {};
    \node[below=of B3] (B4) {\lstinline|FD c ⊗ ((FD c ⊗ FD (τ c)) ⊗ FD (τ c))|};
    \node[right=of B1] (C1) {\lstinline|FD (τ c) ⊗ FD c|};
    \node[below=of C1] (C2) {\lstinline|FD c ⊗ FD (τ c)|};
    \node[below=of C2] (C3) {\lstinline|FD c ⊗ (𝟙_(Rep k G) ⊗ FD (τ c))|};

    \draw[->] (B1) -- (A1);
    \draw[->] (B1) -- (C1);
    \draw[->] (A1) -- (A2);
    \draw[->] (A2) -- (A3);
    \draw[->] (A3) -- (B4);
    \draw[->] (B4) -- (C3);
    \draw[->] (C3) -- (C2);
    \draw[->] (C2) -- (C1);

    \node at ($(B1)!0.3!(A1)$) [above] {\lstinline|(ρ_)⁻¹|};
    \node at ($(B1)!0.45!(C1)$) [above] {\lstinline|unit c|};
    \node at ($(A1)!0.5!(A2)$) [left] {\lstinline|metric c ⊗ metric (τ c)|};
    \node at ($(A2)!0.5!(A3)$) [left] {\lstinline|α_|};
    \node at ($(A3)!0.5!(B4)$) [left, xshift=-5mm] {\lstinline|𝟙 ⊗ (α_)⁻¹|};
    \node at ($(B4)!0.5!(C3)$) [right, xshift=5mm] {\lstinline|𝟙 ⊗ (contr c ⊗ 𝟙)|};
    \node at ($(C3)!0.5!(C2)$) [right] {\lstinline|𝟙 ⊗ λ_|};
    \node at ($(C2)!0.5!(C1)$) [right] {\lstinline|β_|};
  \end{tikzpicture}
\end{equation*}
This roughly says that contracting a metric of color \myinline|c| with that of color 
\myinline|τ c| gives the unit of color \myinline|c|.

For complex Lorentz tensors, the contraction is defined through the dot product,
e.g., the contraction of $\psi^\mu$ and $\phi_\mu$ is via dot product of the underlying vectors.
The metric is defined through the Minkowski metric and the metric tensors, e.g., $\epsilon^{\alpha \dot \alpha}$
for Weyl fermions. Lastly, the units are defined through identity matrices. 

\subsection{Tensors} 

Given any type \myinline|C|, we define the category \myinline|OverColor C| as follows. 
Objects are functions \myinline|f : X → C| for some type \myinline|X|. A morphism from
\myinline|f : X → C| to \myinline|g : Y → C| is a bijection
\myinline|φ : X → Y| such that \myinline|f = g ∘ φ|. This category is equivalent to the core of the
category of types sliced over \myinline|C|. 

The category \myinline|OverColor C| carries a symmetric monoidal structure,
which we will denote \myinline|⊗|. The structure such that \myinline|f ⊗ g| for objects \myinline|f| and \myinline|g|
is the induced map \myinline|X ⊕ Y → C| where \myinline|⊕| denotes the disjoint union of types.
In Lean this is denoted \myinline|Sum.elim f g|. 

For a given tensor species \myinline|S|, the functor \myinline|S.FD| can be lifted 
to a symmetric monoidal functor from \myinline|OverColor S.C| to \myinline|Rep k G|.
Here the monoidal structure on \myinline|Rep k G| is the tensor product over \myinline|k|.
This functor takes \myinline|f : X → S.C| to the tensor product over \myinline|k| of all \myinline|FD (f x)|,
\myinline|⨂[k] x, S.FD (f x)|. 
This construction is general and functorial, allowing us to define the functor: 
\begin{code}
def OverColor.lift : (Discrete S.C ⥤ Rep S.k S.G) ⥤ BraidedFunctor (OverColor S.C) (Rep S.k S.G) where ...
\end{code}
from functors from \myinline|Discrete S.C| to \myinline|Rep S.k S.G|
to symmetric monoidal functors (or braided functors)  from \myinline|OverColor S.C| to \myinline|Rep S.k S.G|.

We denote the lift of \myinline|S.FD| by \myinline|S.F|, which is defined through
\begin{code}
def F  (S : TensorSpecies) : BraidedFunctor (OverColor S.C) (Rep S.k S.G) := 
  (OverColor.lift).obj S.FD
\end{code}

We can think of an object \myinline|f : X → S.C| of \myinline|OverColor S.C| as a type of indices \myinline|X|,
and a specification of what color or representation each index is associated to. 
For example, for the tensor $\phi^{\mu}_{\phantom{\mu}\nu}$ would have \myinline|X| as the type of indices which, since there are two of them, 
is \myinline|Fin 2|, and \myinline|f| as the function which assigns to each index the color of the index, 
so \myinline|f 0| would be \myinline|Color.up|, and \myinline|f 1| would be \myinline|Color.down|. 
We can apply the functor \myinline|S.F| to \myinline|f : X → S.C| in Lean as follows 
\myinline|S.F.obj (OverColor.mk f)|, which we will abbreviate to \myinline|S.F f|.
This representation, \myinline|S.F f|, is the tensor product of each of representations \myinline|S.FD (f x)| for \myinline|x : X|. 
In our example this is equivalent to \myinline|Lorentz.complexContr ⊕ Lorentz.complexCo|. Vectors of the form
\myinline|v : S.F f| can be thought of as tensors with indices indexed by \myinline|X| of color \myinline|C|. 

With this in mind, we define a general tensor of a species \myinline|S| as a vector in a representation 
\myinline|S.F f| for some \myinline|f : OverColor S.C|. 

In physics, we typically focus on objects \myinline|f : X → S.C| of \myinline|OverColor S.C|  where \myinline|X| is a 
finite type of the form \myinline|Fin n| for some \myinline|n : ℕ|. In most of what follows, 
we will restrict to these objects.

\subsection{Tensor Trees and their map to tensors}

Tensor trees are trees with a node for each of the basic operations one can perform on a tensor. 
Namely, tensor trees have nodes for addition of tensors, permutation of tensor indices, negation of tensors, 
  scalar multiplication of tensors, group action on a tensor, tensor product of tensors, contraction of tensor indices,
  and evaluation of tensor indices.
They also have nodes for tensors themselves.

Given a species \myinline|S|, we have a type of tensor tree for each map of the form
\myinline|c : Fin n → S.C| in \myinline|OverColor S.C|. 
This restriction to \myinline|Fin n| is done for convenience.

Tensor trees are defined inductively through a number of constructors: 
\begin{codeLong}
inductive TensorTree (S : TensorSpecies) : {n : ℕ} → (Fin n → S.C) → Type where
  /-- A general tensor node. -/
  | tensorNode {n : ℕ} {c : Fin n → S.C} (T : S.F.obj (OverColor.mk c)) : TensorTree S c
  /-- A node corresponding to the scalar multiple of a tensor by a element of the field. -/
  | smul {n : ℕ} {c : Fin n → S.C} : S.k → TensorTree S c → TensorTree S c
  /-- A node corresponding to negation of a tensor. -/
  | neg {n : ℕ} {c : Fin n → S.C} : TensorTree S c → TensorTree S c
  /-- A node corresponding to the addition of two tensors. -/
  | add {n : ℕ} {c : Fin n → S.C} : TensorTree S c → TensorTree S c → TensorTree S c
  /-- A node corresponding to the action of a group element on a tensor. -/
  | action {n : ℕ} {c : Fin n → S.C} : S.G → TensorTree S c → TensorTree S c
  /-- A node corresponding to the permutation of indices of a tensor. -/
  | perm {n m : ℕ} {c : Fin n → S.C} {c1 : Fin m → S.C}
      (σ : (OverColor.mk c) ⟶ (OverColor.mk c1)) (t : TensorTree S c) : TensorTree S c1
  /-- A node corresponding to the product of two tensors. -/
  | prod {n m : ℕ} {c : Fin n → S.C} {c1 : Fin m → S.C}
    (t : TensorTree S c) (t1 : TensorTree S c1) : TensorTree S (Sum.elim c c1 ∘ finSumFinEquiv.symm)
  /-- A node corresponding to the contraction of indices of a tensor. -/
  | contr {n : ℕ} {c : Fin n.succ.succ → S.C} : (i : Fin n.succ.succ) →
    (j : Fin n.succ) → (h : c (i.succAbove j) = S.τ (c i)) → TensorTree S c →
    TensorTree S (c ∘ Fin.succAbove i ∘ Fin.succAbove j)
  /-- A node corresponding to the evaluation of an index of a tensor. -/
  | eval {n : ℕ} {c : Fin n.succ → S.C} : (i : Fin n.succ) → (x : ℕ) → TensorTree S c →
    TensorTree S (c ∘ Fin.succAbove i)
\end{codeLong}
Each constructor here, e.g., \myinline|tensorNode|, \myinline|smul|, \myinline|neg|, etc., can 
be thought of as forming a different type of node in a tensor tree.

Since the interpretation of each of the constructors is down to how we turn them into a tensor,
we discuss this before outlining each of the constructors in turn. The process 
of going from a tensor tree to a tensor is proscribed by a function
\myinline|TensorTree S c  → S.F c|, which is defined recursively as follows: 
\begin{code}
/-- The underlying tensor a tensor tree corresponds to. -/
def TensorTree.tensor {n : ℕ} {c : Fin n → S.C} : TensorTree S c → S.F.obj (OverColor.mk c) := fun
  | tensorNode t => t
  | smul a t => a • t.tensor
  | neg t => - t.tensor
  | add t1 t2 => t1.tensor + t2.tensor
  | action g t => (S.F.obj (OverColor.mk _)).ρ g t.tensor
  | perm σ t => (S.F.map σ).hom t.tensor
  | prod t1 t2 => (S.F.map (OverColor.equivToIso finSumFinEquiv).hom).hom
    ((S.F.μ _ _).hom (t1.tensor ⊗ₜ t2.tensor))
  | contr i j h t => (S.contrMap _ i j h).hom t.tensor
  | eval i e t => (S.evalMap i (Fin.ofNat' _ e)) t.tensor
\end{code}

Let us now discuss each of the constructors in turn.  

\paragraph{tensorNode:} The constructor \myinline|tensorNode|
creates a tensor tree based on \myinline|c| from a tensor \myinline|t| in \myinline|S.F c|. 
This tensor tree consists of a single node that directly represents the tensor. 
Since all other tensor tree constructors require an existing tensor tree as input, 
\myinline|tensorNode| serves as the foundational base case for building more complex trees. 
As a diagram such a tree is:
\tensorTree{
  \begin{tikzpicture}
    \node[draw=black] (A) at (0,0) {\lstinline|t|};
  \end{tikzpicture} 
}
Naturally, the tensor associated with this node is exactly the tensor provided during construction, 
as can be seen in \myinline|TensorTree.tensor|.

\paragraph{smul:} The constructor smul takes a scalar \myinline|a| and an existing tensor tree 
\myinline|t| based on \myinline|c| and constructs a new tensor tree also based on \myinline|c|. Conceptually, this new tree has a root node labeled 
\myinline|smul a|, with the tensor tree 
\myinline|t| as its child. As a diagram such a tree is:
\tensorTree{
  \begin{tikzpicture}
    \node[draw=black] (A) at (0,0) {\lstinline|smul a|};
    \node (B) at (0,-1) {\lstinline|t|};
    \draw[->] (A) -- (B);
  \end{tikzpicture} 
}
The tensor associated with this new tree is obtained by multiplying the tensor associated with 
\myinline|t| by the scalar \myinline|a|.

\paragraph{neg:} The constructor \myinline|neg| takes an existing tensor tree \myinline|t|
 based on \myinline|c| and constructs a new tensor tree also based on \myinline|c|.
This new tree has a root node labeled \myinline|neg|, with the tensor tree \myinline|t| as its child.
As a diagram, we have:
\tensorTree{
  \begin{tikzpicture}
    \node[draw=black] (A) at (0,0) {\lstinline|neg|};
    \node (B) at (0,-1) {\lstinline|t|};
    \draw[->] (A) -- (B);
  \end{tikzpicture} 
}
The tensor associated with this new tree is obtained by negating the tensor associated with \myinline|t|.

\paragraph{add:} 
The constructor \myinline|add| takes two existing tensor trees \myinline|t1| and \myinline|t2|, based on the same \myinline|c|, and 
constructs a new tensor tree. This new tree has a root node labeled \myinline|add|, with the tensor trees \myinline|t1| and \myinline|t2| as its children.
As a diagram, this corresponds to:
\tensorTree{
  \begin{tikzpicture}
    \node[draw=black] (A) at (0,0) {\lstinline|add|};
    \node (B) at (-1,-1) {\lstinline|t1|};
    \node (C) at (1,-1) {\lstinline|t2|};
    \draw[->] (A) -- (B);
    \draw[->] (A) -- (C);
  \end{tikzpicture} 
}
The tensor associated with this new tree is obtained by adding the tensors associated with \myinline|t1| and \myinline|t2|.

\paragraph{action:}
The constructor \myinline|action| takes a group element \myinline|g| of \myinline|S.G|, an existing tensor tree \myinline|t| based on \myinline|c|, and constructs a new tensor tree also based on \myinline|c|.
This new tree has a root node labeled \myinline|action g|, with the tensor tree \myinline|t| as its child.
As a diagram:
\tensorTree{
  \begin{tikzpicture}
    \node[draw=black] (A) at (0,0) {\lstinline|action g|};
    \node (B) at (0,-1) {\lstinline|t|};
    \draw[->] (A) -- (B);
  \end{tikzpicture} 
}
The tensor associated with this new tree is obtained by acting on the tensor associated with \myinline|t| with the group element \myinline|g|.

\paragraph{perm:}
The constructor \myinline|perm| takes a morphism \myinline|σ| from \myinline|c| to \myinline|c1| in \myinline|OverColor S.C| and 
  an existing tensor tree \myinline|t| based on \myinline|c|, and constructs a new tensor tree based on \myinline|c1|.
This new tree has a root node labeled \myinline|perm σ|, with the tensor tree \myinline|t| as its child.
As a diagram, this is given by: 
\tensorTree{
  \begin{tikzpicture}
    \node[draw=black] (A) at (0,0) {\lstinline|perm σ|};
    \node (B) at (0,-1) {\lstinline|t|};
    \draw[->] (A) -- (B);
  \end{tikzpicture} 
}
The tensor associated with this new tree is obtained by  applying the image of the morphism \myinline|σ| under the functor \myinline|S.F|
to the tensor associated with \myinline|t|.

\paragraph{prod:}
The constructor \myinline|prod| takes two existing tensor trees \myinline|t| based on \myinline|c| and \myinline|t1| based on \myinline|c1| and constructs a new tensor tree
based on \myinline|Sum.elim c c1 ∘ finSumFinEquiv.symm| which is the map from \myinline|Fin (n + n1)| 
acting via \myinline|c i| on \myinline|0 ≤ i ≤ n -1|  and via \myinline|c1 (i - n)| on \myinline|n ≤ i ≤ (n + n1) - 1|.
This new tree has a root node labeled \myinline|prod|, with the tensor trees \myinline|t| and \myinline|t1| as its children.
As a diagram, this is given by:
\tensorTree{
  \begin{tikzpicture}
    \node[draw=black] (A) at (0,0) {\lstinline|prod|};
    \node (B) at (-1,-1) {\lstinline|t|};
    \node (C) at (1,-1) {\lstinline|t1|};
    \draw[->] (A) -- (B);
    \draw[->] (A) -- (C);
  \end{tikzpicture} 
}
The tensor associated with this new tree is obtained by taking the tensor product of the tensors associated with \myinline|t| and \myinline|t1|, 
giving a vector in \myinline|S.F c ⊗ S.F c1|, using the tensorator of \myinline|S.F| (the natural isomorphism 
 between \myinline|S.F _ ⊗ S.F _| and \myinline|S.F (_ ⊗  _)|) to map this vector into 
a vector in \myinline|S.F.obj (OverColor.mk c ⊗ OverColor.mk c1)|, and finally using an isomorphism between \myinline|OverColor.mk c ⊗ OverColor.mk c1| and 
\myinline|OverColor.mk (Sum.elim c c1 ∘ finSumFinEquiv.symm)| to map this vector into
 \myinline|S.F (Sum.elim c c1 ∘ finSumFinEquiv.symm)|.

\paragraph{contr:}
The constructor \myinline|contr| firstly, takes an existing tensor tree \myinline|t| based on a 
\myinline|c : Fin n.succ.succ → S.C|. Here \myinline|n.succ.succ| is $n + 1 + 1$ with \myinline|succ| meaning 
the successor of a natural number. 
It also takes,
an \myinline|i| of type \myinline|Fin n.succ.succ|, \myinline|j| of type \myinline|Fin n.succ|
and a proof that \myinline|c (i.succAbove j) = S.τ (c i)|, where \myinline|i.succAbove| is the map from \myinline|Fin n.succ| to \myinline|Fin n.succ.succ| with a hole at \myinline|i|. 
The proof says that the color of the index \myinline|i.succAbove j| is the dual of the color of the index \myinline|i|, and thus these 
two indices can be contracted.
Note that we use a \myinline|j| in \myinline|Fin n.succ| and \myinline|i.succAbove j| as the index to be contracted, instead 
of another index in  \myinline|Fin n.succ.succ| to ensure the two indices to be contracted are not the same. 
The constructor outputs a new tensor tree based on \myinline|c ∘ i.succAbove ∘ j.succAbove|. 
This new tree has a root node labeled \myinline|contr i j|, with the tensor tree \myinline|t| as its child.
As a diagram:
\tensorTree{
  \begin{tikzpicture}
    \node[draw=black] (A) at (0,0) {\lstinline|contr i j|};
    \node (B) at (0,-1) {\lstinline|t|};
    \draw[->] (A) -- (B);
  \end{tikzpicture} 
}
The tensor associated with the new tensor tree is constructed as follows. 
We start with a tensor in \myinline|S.F c|.
This is mapped into a vector in \myinline|(S.FD (c i) ⊗ S.FD (S.τ (c i))) ⊗ S.F (c ∘ Fin.succAbove i ∘ Fin.succAbove j)| via an equivalence from 
\myinline|S.F c|. The equivalence is constructed using an equivalence 
in \myinline|OverColor C| to extract \myinline|i| and \myinline|i.succAbove j| from \myinline|c|,
then using the tensorator of \myinline|S.F|, and the fact that \myinline|c (i.succAbove j) = S.τ (c i)|. 
Using the contraction for \myinline|c i|, we then get a vector in 
\myinline|𝟙 ⊗ S.F (c ∘ Fin.succAbove i ∘ Fin.succAbove j)| which is then 
mapped to a tensor in \myinline|S.F (c ∘ Fin.succAbove i ∘ Fin.succAbove j)|
using the left-unitor of \myinline|Rep S.G S.k|.
This is all contained within the \myinline|S.contrMap| appearing in the definition of
function \myinline|TensorTree.tensor|.

\paragraph{eval:}
The final constructor \myinline|eval| takes an existing tensor tree \myinline|t| based on a \myinline|c : Fin n.succ → S.C|,
an \myinline|i| of type \myinline|Fin n.succ|, and a natural number $x$.  
The constructor outputs a new tensor tree based on \myinline|c ∘ Fin.succAbove i|.
This new tree has a root node labeled \myinline|eval i x|, with the tensor tree \myinline|t| as its child.
As a diagram, this corresponds to:
\tensorTree{
  \begin{tikzpicture}
    \node[draw=black] (A) at (0,0) {\lstinline|eval i x|};
    \node (B) at (0,-1) {\lstinline|t|};
    \draw[->] (A) -- (B);
  \end{tikzpicture} 
}
The tensor associated with the new tensor tree is constructed as follows.
We start with a tensor in \myinline|S.F c|.
This is mapped into a vector in \myinline|S.FD (c i) ⊗ S.F (c ∘ Fin.succAbove i)| via an equivalence from
\myinline|S.F c|. The equivalence is constructed using an equivalence
in \myinline|OverColor C| to extract \myinline|i| from \myinline|c|, and
then using the tensorator of \myinline|S.F|.
Using the evaluation for \myinline|c i| at the basis element indicated by \myinline|x| (if \myinline|x| is too big a natural number for the number of basis elements it defaults to \myinline|0|), we then get a vector in
\myinline|𝟙 ⊗ S.F (c ∘ Fin.succAbove i)|.
This mapping should be thought of as occurring in the  category of modules over \myinline|S.k|, rather than in 
\myinline|Rep S.G S.k|, since the evaluation is not invariant under the group action.
We then finally map into \myinline|S.F (c ∘ Fin.succAbove i)| 
using the left-unitor. This is all contained within the \myinline|S.evalMap| appearing in the definition of
function \myinline|TensorTree.tensor|.

The main reason tensor trees are easy to work with is the following. 
Define a subtree of a tensor trees 
to be a node and all child nodes of that node. 
If \myinline|t| is a tensor tree and \myinline|s| a subtree of \myinline|t|, we can replace  \myinline|s| in \myinline|t| 
with another tensor tree \myinline|s'|
to get a new overall tensor tree \myinline|t'|. If \myinline|s| and \myinline|s'| have the same underlying tensor, 
then \myinline|t| and \myinline|t'| will also. 

In Lean this property manifests in a series of lemmas. For instance,
for the \myinline|contr| constructor we have the lemma: 
\begin{code}
lemma contr_tensor_eq {n : ℕ} {c : Fin n.succ.succ → S.C} {T1 T2 : TensorTree S c}
    (h : T1.tensor = T2.tensor) {i : Fin n.succ.succ} {j : Fin n.succ}
    {h' : c (i.succAbove j) = S.τ (c i)} :
    (contr i j h' T1).tensor = (contr i j h' T2).tensor := by
  simp only [Nat.succ_eq_add_one, contr_tensor]
  rw [h]
\end{code}
These lemmas allow us to navigate to certain places in tensor trees and replace 
subtrees with other subtrees. We will see this used extensively in the examples in Section~\ref{sec:examples}. 

\subsection{Syntax and their map to tensor trees}

Syntax allows index notation in Lean code to look similar to 
pen-and-paper index notation. 
The syntax is turned into a tensor tree through a process called elaboration. 
Although, elaboration is not formally verified in Lean, the tensor tree it outputs is.

Instead of delving into the finer details of this process, 
we give illustrative examples. 

In what follows we will assume that \myinline|T|, \myinline|T1|, etc. are tensors defined as 
elements of \myinline|S.F c|, \myinline|S.F c1|, etc for 
some tensor species \myinline|S| and some \myinline|c : Fin n → S.C|, \myinline|c1 : Fin n1 → S.C|, etc.
for which the expressions below make sense.

The syntax allows us to write the following:
\syntaxElab{\lstinline!\{T | μ ν\}ᵀ!}{{\lstinline!tensorNode T!}}
for a tensor node. Here the \myinline|μ| and \myinline|ν| are free variables and it does not 
matter what we call them - Lean will elaborate the expression in the same way.
The elaborator also knows how many indices to expect for a tensor \myinline|T| and will raise an error if
the wrong number are given. The \myinline|{_}ᵀ| notation is used to tell Lean that the syntax
is to be treated as a tensor expression. Throughout this section we will use the two-sided boxes 
given above, which denote the syntax on the left and the expression it is elaborated to on the right. 

We can write, for example:
\syntaxElab{\lstinline!\{T | μ ν\}ᵀ.tensor!}{{\lstinline!(tensorNode T).tensor!}}
to get the underlying tensor. We get this notation from the way \myinline|TensorTree.tensor|
is defined with the prefix \myinline|TensorTree.|.

Note that we do not have indices which are upper or lower as one would expect from pen-and-paper notation
 (e.g., $\eta^\mu_{\phantom{\mu}\nu}$).
There is one primary reason for this; whether an index is upper or lower does not carry any information, 
since this information comes from the tensor itself. Also, for something like complex Lorentz tensors, 
there are three different types of upper-index, so such notation would be complicated. 

If we want to evaluate an index we can put an explicit index in place of $\mu$ or $\nu$ above, 
for example:
\syntaxElab{\lstinline!\{T | 1 ν\}ᵀ!}{\lstinline!eval 0 1 (tensorNode T)!}

The syntax and elaboration for negation, scalar multiplication and the group action 
are fairly similar. For negation we have:
\syntaxElab{\lstinline!\{- T | μ ν\}ᵀ!}{\lstinline!neg (tensorNode T)!}
For scalar multiplication by $a \in k$ we have:
\syntaxElab{\lstinline!\{a •ₜ T | μ ν\}ᵀ!}{\lstinline!smul a (tensorNode T)!}
For the group action of $g\in G$ on a tensor $T$ we have:
\syntaxElab{\lstinline!\{g •ₐ T | μ ν\}ᵀ!}{\lstinline!action g (tensorNode T)!}
The product of two tensors is also fairly similar, with us having: 
\syntaxElab{\lstinline!\{T | μ ν ⊗ T2 | σ\}ᵀ!}{\lstinline!prod (tensorNode T) (tensorNode T2)!}

The syntax for contraction is done by pairing indices: 
\syntaxElab{\lstinline!\{T | μ ν ⊗ T2 | ν σ\}ᵀ!}{\lstinline!contr 1 1 rfl (prod (tensorNode T) (tensorNode T2))!}
On the right-hand side the first argument (\myinline!1!) of \myinline!contr! is the index of the first \myinline|ν| on the left-hand side, 
the second argument  (also \myinline!1!) is the index of the second \myinline|ν| after the first is removed. The \myinline!rfl! is a proof that the
colors of the two contracted indices are actually dual to one another. If they are not, this proof will 
fail and the elaborator will complain. It will also complain if more than two indices are trying 
to be contracted, although this depends on where exactly the indices sit in the expression, for example 
\syntaxElab{\lstinline!\{T | μ ν ⊗ T2 | ν ν\}ᵀ!}{\lstinline!(prod (tensorNode T) (contr 0 0 rfl (tensorNode T2)))!}
works fine because the inner contraction is computed before the product. 

We now turn to addition. Our syntax allows for, e.g., \myinline!{T | μ ν + T2 | μ ν}ᵀ! and also 
\myinline!{T | μ ν + T2 | ν μ}ᵀ!, provided that the indices are of the correct color (which Lean will check). 
The elaborator handles both these cases and generalizations thereof by adding a permutation node. Thus we have:
\syntaxElab{\lstinline!\{T | μ ν + T2 | μ ν\}ᵀ!}{\lstinline!add (tensorNode T) (perm _ (tensorNode T2))!}
where the \myinline!_! is a placeholder for the permutation, something we will use frequently in what follows. For the case above the permutation will be the identity, but for: 
\syntaxElab{\lstinline!\{T | μ ν + T2 | ν μ\}ᵀ!}{\lstinline!add (tensorNode T) (perm _ (tensorNode T2))!}
it will be the permutation for the two indices. 

Despite not forming part of a node in our tensor tree, we also give syntax for equality. 
This is done in a very similar way to addition, with the inclusion of a permutation node to account for,
e.g., expressions like $T_{\mu \nu} = T_{\nu \mu}$. In particular, we have:
\syntaxElab{\lstinline!\{T | μ ν = T2 | ν μ\}ᵀ!}{\lstinline!(tensorNode T).tensor = (perm _ (tensorNode T2)).tensor!}
Note that in the elaborated expression we ask for equality of
tensors through \myinline!.tensor!. Tensors are, after all, the objects we care about.

With this syntax we can write complicated tensor expressions in a way close to pen-and-paper index notation. 
We will see examples of this in the next section.

\section{Examples} \label{sec:examples}

We give two examples in this section. The first example is a simple theorem involving index notation and 
tensor trees. We will demonstrate, in rather explicit detail, how we can manipulate tensor trees to solve 
such theorems. 
The second example we give will show a number of definitions related to Pauli matrices and bispinors in HepLean concerning index notation.
Here we won't give as much detail, the point being to show the reader the broad use of our 
construction. 

\subsection{Example 1: Symmetric and antisymmetric tensor} \label{sec:exampleSymmAntiSymm}
If $A^{\mu \nu}$ is an antisymmetric tensor and $S_{\mu \nu}$ and $S$ is a symmetric tensor, then
it is true that $A^{\mu \nu} S_{\mu \nu} = - A^{\mu \nu} S_{\mu \nu}$. In Lean this result, and 
its proof are written as follows: 
\begin{codeLong}
lemma antiSymm_contr_symm 
    {A : complexLorentzTensor.F.obj (OverColor.mk ![Color.up, Color.up])}
    {S : complexLorentzTensor.F.obj (OverColor.mk ![Color.down, Color.down])}
    (hA : {A | μ ν = - (A | ν μ)}ᵀ) (hs : {S | μ ν = S | ν μ}ᵀ) :
    {A | μ ν ⊗ S | μ ν = - A | μ ν ⊗ S | μ ν}ᵀ := by
  conv =>
    lhs
    rw [contr_tensor_eq <| contr_tensor_eq <| prod_tensor_eq_fst <| hA]
    rw [contr_tensor_eq <| contr_tensor_eq <| prod_tensor_eq_snd <| hs]
    rw [contr_tensor_eq <| contr_tensor_eq <| prod_perm_left _ _ _ _]
    rw [contr_tensor_eq <| contr_tensor_eq <| perm_tensor_eq <| prod_perm_right _ _ _ _]
    rw [contr_tensor_eq <| contr_tensor_eq <| perm_perm _ _ _]
    rw [contr_tensor_eq <| perm_contr_congr 1 2]
    rw [perm_contr_congr 0 0]
    rw [perm_tensor_eq <| contr_contr _ _ _]
    rw [perm_perm]
    rw [perm_tensor_eq <| contr_tensor_eq <| contr_tensor_eq <| neg_fst_prod _ _]
    rw [perm_tensor_eq <| contr_tensor_eq <| neg_contr _]
    rw [perm_tensor_eq <| neg_contr _]
  apply perm_congr _ rfl
  decide
\end{codeLong}
Let us break this down. The statements 
\begin{code} 
{A : complexLorentzTensor.F.obj (OverColor.mk ![Color.up, Color.up])}
{S : complexLorentzTensor.F.obj (OverColor.mk ![Color.down, Color.down])}
\end{code}
are simply defining $A$ and $S$ to be tensors of type $A^{\mu \nu}$ and $S_{\mu \nu}$ respectively.
Here \myinline|![Color.up, Color.up]| is shorthand for the map \myinline|Fin 2 → complexLorentzTensor.C| that sends
$0$ to \myinline|Color.up| and $1$ to \myinline|Color.up|.

The parameter \myinline|hA| is stating that $A$ is antisymmetric. Expanded in terms of tree diagrams 
we have
\proofstep{\lstinline!hA : \{A | μ ν = - (A | ν μ)\}ᵀ!}{Description: The tensor $A$ 
  is antisymmetric.}{
 \begin{tikzpicture}
    \node[draw=black] (A) at (-2,-1) {A};
    \node[draw=black] (D1) at (0,0) {perm \_};
    \node[draw=black] (E1) at (0,-1) {neg};
    \node[draw=black] (F1) at (0,-2) {A};
    \node (eq) at (-1, -1) {$=$};
    \path [->] (D1) edge (E1);
    \path [->] (E1) edge (F1);
  \end{tikzpicture} 
}
In the tensor tree diagram on the right-hand side we implicitly mean 
equality of the underlying tensors of the given trees. 
This will be left implicit throughout.  

Similarly, the parameter \myinline|hs| is stating that $S$ is symmetric. Expanded in terms of tree diagrams
\proofstep{\lstinline!hS : \{S | μ ν = S | ν μ\}ᵀ!}{Description: The tensor \myinline|S|
  is symmetric.}{
 \begin{tikzpicture}
    \node[draw=black] (A) at (-2,-0.5) {S};
    \node[draw=black] (D1) at (0,0) {perm \_};
    \node[draw=black] (F1) at (0,-1) {S};
    \node (eq) at (-1, -0.5) {$=$};
    \path [->] (D1) edge (F1);
  \end{tikzpicture} 
}

The line \myinline!{A | μ ν ⊗ S | μ ν = - A | μ ν ⊗ S | μ ν}ᵀ! is the statement we are trying to prove. 
In terms of tree diagrams it says that:
\begin{center}
  \fcolorbox{mycolor}{white}{
\begin{tikzpicture}
  \node (eq) at (2,-1) {$=$};
  \node[draw=black] (A) at (0,0) {contr 0 0};
  \node[draw=black] (B) at (0,-1) {contr 0 1};
  \node[draw=black] (C) at (0,-2) {prod};
  \node[draw=black] (D1) at (-1,-3) {A};
  \node[draw=black] (D2) at (1,-3) {S};
  \path [->] (A) edge (B);
  \path [->] (B) edge (C);
  \path [->] (C) edge (D1);
  \path [->] (C) edge (D2);
  \node[draw=black] (P') at (4,1) {perm $\_$};
  \node[draw=black] (N') at (4,0) {neg};
  \node[draw=black] (A') at (4,-1) {contr 0 0};
  \node[draw=black] (B') at (4,-2) {contr 0 1};
  \node[draw=black] (C') at (4,-3) {prod};
  \node[draw=black] (D1') at (3,-4) {A};
  \node[draw=black] (D2') at (5,-4) {S};
  \path [->] (P') edge (N');
  \path [->] (N') edge (A');
  \path [->] (A') edge (B');
  \path [->] (B') edge (C');
  \path [->] (C') edge (D1');
  \path [->] (C') edge (D2');
\end{tikzpicture}}
\end{center}
The perm here actually does nothing, but is included by Lean. 

The lines of the proof in the \myinline|conv| block are manipulations of the tensor tree on the 
LHS of the equation. The \myinline|rw| tactic is used to rewrite the tensor tree using the various lemmas. 
We work through each step in turn. 
\proofstep{\lstinline!rw [contr_tensor_eq <| contr_tensor_eq <| prod_tensor_eq_fst <| hA]!}{
  Description: Here
  \myinline|contr_tensor_eq| and \myinline|prod_tensor_eq_fst| navigate to the correct place in the tensor tree, 
  whilst 
  \myinline|hA| replaces the node \myinline|A| with the RHS of \myinline|hA|.
}{
  \begin{tikzpicture}
    \node[draw=black] (A) at (0,0) {contr 0 0};
    \node[draw=black] (B) at (0,-1) {contr 0 1};
    \node[draw=black] (C) at (0,-2) {prod};
    \node[draw=mydarkcolor] (D1) at (-1,-3) {perm \_};
    \node[draw=mydarkcolor] (E1) at (-1,-4) {neg};
    \node[draw=mydarkcolor] (F1) at (-1,-5) {A};
    \node[draw=black] (D2) at (1,-3) {S};
    \path [->] (A) edge (B);
    \path [->] (B) edge (C);
    \path [->] (C) edge (D1);
    \path [->, color = mydarkcolor] (D1) edge (E1);
    \path [->, color = mydarkcolor] (E1) edge (F1);
    \path [->] (C) edge (D2);
  \end{tikzpicture}
}

\proofstep{\lstinline!rw [contr_tensor_eq <| contr_tensor_eq <| prod_tensor_eq_snd <| hs]!}{
  Description: 
  Here
  \myinline|contr_tensor_eq| and \myinline|prod_tensor_eq_fst| navigate to the correct place in the tensor tree, 
  whilst 
  \myinline|hS| replaces the node \myinline|S| with the RHS of \myinline|hS|.
}{
  \begin{tikzpicture}
    \node[draw=black] (A) at (0,0) {contr 0 0};
    \node[draw=black] (B) at (0,-1) {contr 0 1};
    \node[draw=black] (C) at (0,-2) {prod};
    \node[draw=black] (D1) at (-1,-3) {perm \_};
    \node[draw=black] (E1) at (-1,-4) {neg};
    \node[draw=black] (F1) at (-1,-5) {A};
    \node[draw=mydarkcolor] (D2) at (1,-3) {perm \_};
    \node[draw=mydarkcolor] (F2) at (1,-4) {S};
    \path [->] (A) edge (B);
    \path [->] (B) edge (C);
    \path [->] (C) edge (D1);
    \path [->] (D1) edge (E1);
    \path [->] (E1) edge (F1);
    \path [->] (C) edge (D2);
    \path [->, color=mydarkcolor] (D2) edge (F2);
  \end{tikzpicture}
}

\proofstep{\lstinline!rw [contr_tensor_eq <| contr_tensor_eq <| prod_perm_left _ _ _ _]!}{
  Description: Here \myinline|contr_tensor_eq| navigates to the correct place in the tensor tree,
  whilst \myinline|prod_perm_left| moves the permutation on the left through the product.
}{
  \begin{tikzpicture}
    \node[draw=black] (A) at (0,0) {contr 0 0};
    \node[draw=black] (B) at (0,-1) {contr 0 1};
    \node[draw=mydarkcolor] (C) at (0,-2) {perm \_};
    \node[draw=mydarkcolor] (D) at (0,-3) {prod};
    \node[draw=black] (E1) at (-1,-4) {neg};
    \node[draw=black] (F1) at (-1,-5) {A};
    \node[draw=black] (E2) at (1,-4) {perm \_};
    \node[draw=black] (F2) at (1,-5) {S};
    \path [->] (A) edge (B);
    \path [->] (B) edge (C);
    \path [->, color = mydarkcolor] (C) edge (D);
    \path [->, color = mydarkcolor] (D) edge (E1);
    \path [->] (E1) edge (F1);
    \path [->] (D) edge (E2);
    \path [->] (E2) edge (F2);
  \end{tikzpicture}
}

\proofstep{\lstinline!rw [contr_tensor_eq <| contr_tensor_eq <| perm_tensor_eq <| prod_perm_right _ _ _ _]!}{
  Description: Here \myinline|contr_tensor_eq| and \myinline|perm_tensor_eq| navigate to the correct place in the tensor tree,
  whilst \myinline|prod_perm_right| moves the permutation on the right through the product.
}{
  \begin{tikzpicture}
    \node[draw=black] (A) at (0,0) {contr 0 0};
    \node[draw=black] (B) at (0,-1) {contr 0 1};
    \node[draw=black] (C) at (0,-2) {perm \_};
    \node[draw=mydarkcolor] (D) at (0,-3) {perm \_};
    \node[draw=mydarkcolor] (E) at (0,-4) {prod};
    \node[draw=black] (F1) at (-1,-5) {neg};
    \node[draw=black] (G1) at (-1,-6) {A};
    \node[draw=black] (F2) at (1,-5) {S};
    \path [->] (A) edge (B);
    \path [->] (B) edge (C);
    \path [->] (C) edge (D);
    \path [->, color = mydarkcolor] (D) edge (E);
    \path [->] (E) edge (F1);
    \path [->, color = mydarkcolor] (E) edge (F2);
    \path [->] (F1) edge (G1);
  \end{tikzpicture}
}

\proofstep{\lstinline!rw [contr_tensor_eq <| contr_tensor_eq <| perm_perm _ _ _]!}{
  Description: Here \myinline|contr_tensor_eq| navigates to the correct place in the tensor tree,
  whilst \myinline|perm_perm| uses functoriality to combine the two permutations.
}{
  \begin{tikzpicture}
    \node[draw=black] (A) at (0,0) {contr 0 0};
    \node[draw=black] (B) at (0,-1) {contr 0 1};
    \node[draw=mydarkcolor] (C) at (0,-2) {perm \_};
    \node[draw=black] (D) at (0,-3) {prod};
    \node[draw=black] (E1) at (-1,-4) {neg};
    \node[draw=black] (F1) at (-1,-5) {A};
    \node[draw=black] (E2) at (1,-4) {S};
    \path [->] (A) edge (B);
    \path [->] (B) edge (C);
    \path [->] (C) edge (D);
    \path [->] (D) edge (E1);
    \path [->] (D) edge (E2);
    \path [->] (E1) edge (F1);
  \end{tikzpicture}
}

\proofstep{\lstinline!rw [contr_tensor_eq <| perm_contr_congr 1 2]!}{
  Description: Here \myinline|contr_tensor_eq| navigates to the correct place in the tensor tree,
  whilst \myinline|perm_contr_congr| moves the permutation through the contraction, and simplifies the contraction
  indices to \myinline|1| and \myinline|2| (Lean will check if this is correct).
}{
  \begin{tikzpicture}
    \node[draw=black] (A) at (0,0) {contr 0 0};
    \node[draw=mydarkcolor] (B) at (0,-1) {perm \_};
    \node[draw=mydarkcolor] (C) at (0,-2) {contr 1 2};
    \node[draw=black] (D) at (0,-3) {prod};
    \node[draw=black] (E1) at (-1,-4) {neg};
    \node[draw=black] (F1) at (-1,-5) {A};
    \node[draw=black] (E2) at (1,-4) {S};
    \path [->] (A) edge (B);
    \path [->, color = mydarkcolor] (B) edge (C);
    \path [->, color = mydarkcolor] (C) edge (D);
    \path [->] (D) edge (E1);
    \path [->] (D) edge (E2);
    \path [->] (E1) edge (F1);
  \end{tikzpicture}
}

\proofstep{\lstinline!rw [perm_contr_congr 0 0]!}{
  Description: Here \myinline|perm_contr_congr| moves the permutation through the contraction, and simplifies the contraction
  indices to \myinline|0| and \myinline|0| (Lean will check if this is correct).
}{
  \begin{tikzpicture}
    \node[draw=mydarkcolor] (A) at (0,0) {perm \_};
    \node[draw=mydarkcolor] (B) at (0,-1) {contr 0 0};
    \node[draw=black] (C) at (0,-2) {contr 1 2};
    \node[draw=black] (D) at (0,-3) {prod};
    \node[draw=black] (E1) at (-1,-4) {neg};
    \node[draw=black] (F1) at (-1,-5) {A};
    \node[draw=black] (E2) at (1,-4) {S};
    \path [->, color = mydarkcolor] (A) edge (B);
    \path [->, color = mydarkcolor] (B) edge (C);
    \path [->] (C) edge (D);
    \path [->] (D) edge (E1);
    \path [->] (D) edge (E2);
    \path [->] (E1) edge (F1);
  \end{tikzpicture}
}

\proofstep{\lstinline!rw [perm_tensor_eq <| contr_contr _ _ _]!}{
  Description: Here \myinline|perm_tensor_eq| navigates to the correct place in the tensor tree,
  whilst \myinline|contr_contr| swaps the two contractions, in the meantime inducing 
  a permutation.
}{
  \begin{tikzpicture}
    \node[draw=black] (A) at (0,0) {perm \_};
    \node[draw=mydarkcolor] (A') at (0,-1) {perm \_};
    \node[draw=mydarkcolor] (B) at (0,-2) {contr 0 0};
    \node[draw=mydarkcolor] (C) at (0,-3) {contr 0 1};
    \node[draw=black] (D) at (0,-4) {prod};
    \node[draw=black] (E1) at (-1,-5) {neg};
    \node[draw=black] (F1) at (-1,-6) {A};
    \node[draw=black] (E2) at (1,-5) {S};
    \path [->] (A) edge (A');
    \path [->, color = mydarkcolor] (A') edge (B);
    \path [->, color = mydarkcolor] (B) edge (C);
    \path [->, color = mydarkcolor] (C) edge (D);
    \path [->] (D) edge (E1);
    \path [->] (D) edge (E2);
    \path [->] (E1) edge (F1);
  \end{tikzpicture}
}

\proofstep{\lstinline!rw [perm_perm]!}{
  Description: Here \myinline|perm_perm| uses functoriality to combine the two permutations.
}{
  \begin{tikzpicture}
    \node[draw=mydarkcolor] (A) at (0,0) {perm \_};
    \node[draw=black] (B) at (0,-1) {contr 0 0};
    \node[draw=black] (C) at (0,-2) {contr 0 1};
    \node[draw=black] (D) at (0,-3) {prod};
    \node[draw=black] (E1) at (-1,-4) {neg};
    \node[draw=black] (F1) at (-1,-5) {A};
    \node[draw=black] (E2) at (1,-4) {S};
    \path [->] (A) edge (B);
    \path [->] (B) edge (C);
    \path [->] (C) edge (D);
    \path [->] (D) edge (E1);
    \path [->] (D) edge (E2);
    \path [->] (E1) edge (F1);
  \end{tikzpicture}
}

\proofstep{\lstinline!rw [perm_tensor_eq <| contr_tensor_eq <| contr_tensor_eq <| neg_fst_prod _ _]!}{
  Description: Here \myinline|perm_tensor_eq| and \myinline|contr_tensor_eq| navigate to the correct place in the tensor tree,
  whilst \myinline|neg_fst_prod| moves the negation through the product.
}{
  \begin{tikzpicture}
    \node[draw=black] (A) at (0,0) {perm \_};
    \node[draw=black] (B) at (0,-1) {contr 0 0};
    \node[draw=black] (C) at (0,-2) {contr 0 1};
    \node[draw=mydarkcolor] (D) at (0,-3) {neg};
    \node[draw=mydarkcolor] (E) at (0,-4) {prod};
    \node[draw=black] (F1) at (-1,-5) {A};
    \node[draw=black] (F2) at (1,-5) {S};
    \path [->] (A) edge (B);
    \path [->] (B) edge (C);
    \path [->] (C) edge (D);
    \path [->, color = mydarkcolor] (D) edge (E);
    \path [->, color = mydarkcolor] (E) edge (F1);
    \path [->] (E) edge (F2);
  \end{tikzpicture}
}

\proofstep{\lstinline!rw [perm_tensor_eq <| contr_tensor_eq <| neg_contr _]!}{
  Description: Here \myinline|perm_tensor_eq| and \myinline|contr_tensor_eq| navigate to the correct place in the tensor tree,
  whilst \myinline|neg_contr| moves the negation through the first contraction.
}{
  \begin{tikzpicture}
    \node[draw=black] (A) at (0,0) {perm \_};
    \node[draw=black] (B) at (0,-1) {contr 0 0};
    \node[draw=mydarkcolor] (C) at (0,-2) {neg};
    \node[draw=mydarkcolor] (D) at (0,-3) {contr 0 1};
    \node[draw=black] (E) at (0,-4) {prod};
    \node[draw=black] (F1) at (-1,-5) {A};
    \node[draw=black] (F2) at (1,-5) {S};
    \path [->] (A) edge (B);
    \path [->] (B) edge (C);
    \path [->, color = mydarkcolor] (C) edge (D);
    \path [->, color = mydarkcolor] (D) edge (E);
    \path [->] (E) edge (F1);
    \path [->] (E) edge (F2);
  \end{tikzpicture}
}

\proofstep{\lstinline!rw [perm_tensor_eq <| neg_contr _]!}{
  Description: Here \myinline|perm_tensor_eq| navigates to the correct place in the tensor tree,
  whilst \myinline|neg_contr| moves the negation through the second contraction.
}{
  \begin{tikzpicture}
    \node[draw=black] (A) at (0,0) {perm \_};
    \node[draw=mydarkcolor] (B) at (0,-1) {neg};
    \node[draw=mydarkcolor] (C) at (0,-2) {contr 0 0};
    \node[draw=black] (D) at (0,-3) {contr 0 1};
    \node[draw=black] (E) at (0,-4) {prod};
    \node[draw=black] (F1) at (-1,-5) {A};
    \node[draw=black] (F2) at (1,-5) {S};
    \path [->] (A) edge (B);
    \path [->, color = mydarkcolor] (B) edge (C);
    \path [->, color = mydarkcolor] (C) edge (D);
    \path [->] (D) edge (E);
    \path [->] (E) edge (F1);
    \path [->] (E) edge (F2);
  \end{tikzpicture}
}

The remainder of the proof
\begin{code}
  apply perm_congr _ rfl
  decide
\end{code}
helps Lean to understand that the two sides of the equation are equal.
\subsection{Example 2: Pauli matrices and bispinors} \label{sec:examplePauliBispinor}
Using the formalism we have set up thus far, it is possible to define Pauli matrices and bispinors 
as complex Lorentz tensors.
 
The Pauli matrices appear in HepLean as follows:
\begin{code}
/-- The Pauli matrices as the complex Lorentz tensor `σ^μ^α^{dot β}`. -/
def pauliContr := {PauliMatrix.asConsTensor | ν α β}ᵀ.tensor

/-- The Pauli matrices as the complex Lorentz tensor `σ_μ^α^{dot β}`. -/
def pauliCo := {η' | μ ν ⊗ pauliContr | ν α β}ᵀ.tensor

/-- The Pauli matrices as the complex Lorentz tensor `σ_μ_α_{dot β}`. -/
def pauliCoDown := {pauliCo | μ α β ⊗ εL' | α α' ⊗ εR' | β β'}ᵀ.tensor

/-- The Pauli matrices as the complex Lorentz tensor `σ^μ_α_{dot β}`. -/
def pauliContrDown := {pauliContr | μ α β ⊗ εL' | α α' ⊗ εR' | β β'}ᵀ.tensor
\end{code}
The first of these definitions depends on \myinline|PauliMatrix.asConsTensor| which is defined 
using an explicit basis expansion as a map: 
\begin{code}
  𝟙_ (Rep ℂ SL(2,ℂ)) ⟶ complexContr ⊗ Fermion.leftHanded ⊗ Fermion.rightHanded
\end{code}
which Lean 
knows how to treat as a tensor. Here \myinline|complexContr| is the representation of complex Lorentz vectors 
under $SL(2, \mathbb{C})$ defined above as \myinline|Lorentz.complexContr|.

In these expressions we have the appearance of metrics. 
The metric \myinline|η'| is what is usually denoted 
$\eta_{\mu \nu}$, the metric \myinline|η| is what is usually denoted $\eta^{\mu \nu}$, the metric \myinline|εL'| is what is usually denoted 
$\epsilon_{\alpha \alpha'}$, and the metric \myinline|εR'| is what is usually denoted $\epsilon_{\dot \beta \dot \beta'}$.

With these we can also define bispinors:
\begin{codeLong}
/-- A bispinor `pᵃᵃ` created from a lorentz vector `p^μ`. -/
def contrBispinorUp (p : complexContr) :=
  {pauliCo | μ α β ⊗ p | μ}ᵀ.tensor

/-- A bispinor `pₐₐ` created from a lorentz vector `p^μ`. -/
def contrBispinorDown (p : complexContr) :=
  {εL' | α α' ⊗ εR' | β β' ⊗ contrBispinorUp p | α β}ᵀ.tensor

/-- A bispinor `pᵃᵃ` created from a lorentz vector `p_μ`. -/
def coBispinorUp (p : complexCo) := {pauliContr | μ α β ⊗ p | μ}ᵀ.tensor

/-- A bispinor `pₐₐ` created from a lorentz vector `p_μ`. -/
def coBispinorDown (p : complexCo) :=
  {εL' | α α' ⊗ εR' | β β' ⊗ coBispinorUp p | α β}ᵀ.tensor
\end{codeLong}
Here  \myinline|complexCo| corresponds to the representation of covariant complex Lorentz vectors, 
defined above as \myinline|Lorentz.complexCo|.

Using these definitions we can start to prove results about the pauli matrices and bispinors. 
These proofs essentially rely on the sorts of manipulations in the last section, although in some cases 
we expand tensors in terms of a basis and use rules about how the basis interacts with the operations in a tensor tree. 

As an example, we prove the lemmas
\begin{code}
lemma coBispinorDown_eq_pauliContrDown_contr (p : complexCo) :
  {coBispinorDown p | α β = pauliContrDown | μ α β ⊗ p | μ}ᵀ := by
\end{code}
and 
\begin{code}
/-- The statement that `η_{μν} σ^{μ α dot β} σ^{ν α' dot β'} = 2 ε^{αα'} ε^{dot β dot β'}`. -/
theorem pauliCo_contr_pauliContr :
    {pauliCo | ν α β ⊗ pauliContr | ν α' β' = 2 •ₜ εL | α α' ⊗ εR | β β'}ᵀ := by
\end{code}
We do not give the proofs of these lemmas explicitly. 
The former, however, is a fairly simple application of associativity of the tensor product, and shuffling 
around of the contractions. 

\section{Future work} \label{sec:future}
\begin{figure}
  \centering
  \includegraphics[width=\textwidth]{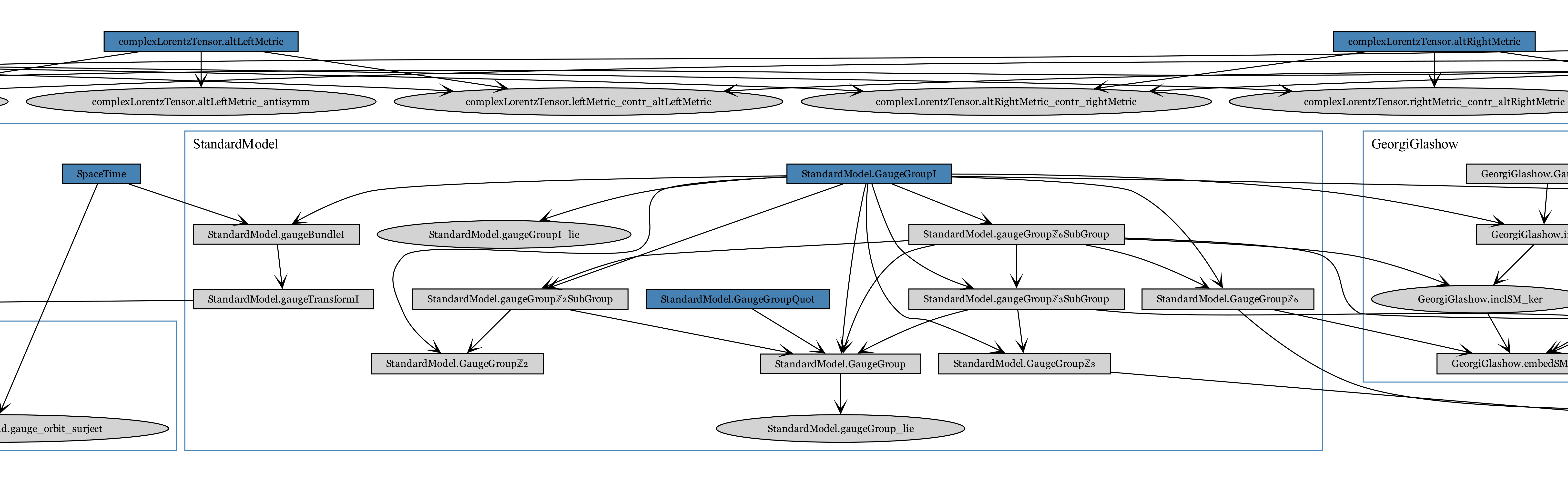}
  \caption{Part of the informal dependency graph in HepLean. Gray nodes indicate informal results, whilst 
  blue nodes indicate results already formalised.}
  \label{fig:informalLemmas}
\end{figure}

Inspired by Lean
blueprints~\cite{leanblueprint}, we have added to HepLean informal lemmas related to index notation and tensors. 
An example of such is: 
\begin{code} 
informal_lemma coBispinorUp_eq_metric_contr_coBispinorDown where
  math :≈ "{coBispinorUp p | α β = εL | α α' ⊗ εR | β β' ⊗ coBispinorDown p | α' β' }ᵀ"
  proof :≈ "Expand `coBispinorDown` and use fact that metrics contract to the identity."
  deps :≈ [``coBispinorUp, ``coBispinorDown, ``leftMetric, ``rightMetric]
\end{code}
These informal lemmas are written in strings, and are not type checked.
They also include dependencies indicating what definitions and lemmas we expect to be used in their statement or proof.
They are intended to be a guide for future formalization efforts, either by humans or 
AI. All the informal lemmas and the related informal definitions which are in HepLean are 
given on the HepLean website in a dependency graph, part of which is shown in Figure \ref{fig:informalLemmas}.

As demonstrated in our earlier examples, manipulating tensor expressions can involve tedious
 calculations, especially when dealing directly with tensor trees. 
In future, we intend to automate many of these routine steps by developing suitable tactics 
within Lean. We are optimistic that the structured nature of tensor trees will lend itself well to 
such automation, thereby streamlining computations and enhancing the efficiency of formal proofs
involving index notation and tensor species.

There are two primary directions in which we can extend the concepts presented in this work. 
First, we could incorporate the spinor-helicity formalism, which is used in the study of scattering 
amplitudes. Second, we could extend our approach to encompass tensor \emph{fields}, their derivatives
etc. We do not anticipate any insurmountable challenges in pursuing these extensions. 
They represent promising avenues for future research and have the potential to significantly enhance
the utility of formal methods in physics.
\section*{Acknowledgments}
I thank Sven Krippendorf, Andreas Schachner, and Tarmo Uustalu for helpful discussions related to this project.
I also thank Tarmo Uustalu for his comments on a draft of this paper.
I also thank Jan Idziak, Thomas Murrills, and Tomas Skrivan for 
dicussions related to index notation near the start of this project. 
I thank members of the Lean Zulip for answering my Lean related questions.
This research is supported by the project ``Icelandic advantage in computer-assisted proof''
of the Collaboration Fund of Iceland's Ministry of Higher Education, Science and Innovation.
\bibliographystyle{unsrturl}
\begin{spacing}{0.5}
\bibliography{MyBib}
\end{spacing}
\end{document}